\documentclass[doublecol,linenumbers]{epl2} 
\usepackage{graphicx}
\usepackage{dcolumn}
\usepackage{bm}
\usepackage{graphicx,amssymb,amsmath,amsthm,amsfonts,epsfig}
\usepackage{subfigure}

\title{Enthalpy, Geometric Volume and Logarithmic correction to Entropy for Van-der-Waals Black Hole }

\author{Parthapratim Pradhan\footnote{pppradhan77@gmail.com}\inst{1}}

\institute{                    
  \inst{1} Department of Physics, Vivekananda Satabarshiki Mahavidyalaya ,
West Midnapur, West Bengal 721513, India \\
}


\abstract{If the negative cosmological constant is treated as a dynamical pressure and if the 
volume be its thermodynamically conjugate variable then the gravitational mass can be expressed as the total gravitational 
enthalpy rather than the energy. Under these circumstances, a new phenomena  emerges in the context of extended phase space 
thermodynamics. We \emph{examine} here these features  for recently discovered Van-der-Waal (VDW) black hole (BH) \cite{mann15}
which is analogous to the VDW fluid. We show that the thermodynamic volume  is \emph{greater} than the naive geometric 
volume. We also show that the \emph{Smarr-Gibbs-Duhem} relation is satisfied for this BH. Furthermore, by computing the 
thermal specific heat we find the local thermodynamic stability criterion for this BH. It has been observed that 
the BH does \emph{not} possess any kind of second order phase transition. This is an interesting feature of 
VDW BH by its own right. Moreover, we also derive \emph{Cosmic-Censorship-Inequality} for this class of BH. 
In addition finally, we compute the \emph{logarithmic correction} to the entropy of this BH due to the quantum  
fluctuations around the thermal equilibrium.}

\begin{document}

\maketitle

\section{Introduction}
There has been a resurgence of interest  recently in the physics of asymptotically anti-de-Sitter (AdS)  spacetime due to
the seminal work of ``Thermodynamics of BHs in ADS Space'' by  Hawking  and Page \cite{hp83} and 
due to the AdS/CFT correspondence \cite{mal98}. The main interest in  AdS case because thermodynamic equilibrium is 
straightforwardly defined and they have a gauge duality description via a dual thermal field theory.

It is also quite interesting that their thermodynamics exhibiting different type of  phase transitions. It was
first observed in case of  Schwarzschild-AdS BH \cite{hp83}, where the first order phase transition 
occurs. On the other hand, classical BH thermodynamics \cite{bk73,bk74,hk75,hk76} states that the mass $M$, surface gravity
$\kappa$, and area $A$ of a BH connected to the thermal energy $U$, temperature $T$, and entropy $S$ via the
following relation (in units in which $G=\hbar=c=k=1$):
\begin{eqnarray}
M=U ,\,\,\, T=\frac{\kappa}{2\pi},\,\,\, S=\frac{A}{4}  ~.\label{mgs}
\end{eqnarray}

In most computations of BH thermodynamics the cosmological constant $\Lambda$ is treated 
as a fixed parameter (possibly zero) but it has been considered as a dynamical variable first in \cite{ht84} and 
it has been further first proposed that it is better to considered as a thermodynamic variable 
rather than a dynamical variable
\cite{kastor09,dolan10,dolan11a,dolan11b,johnson14}. Physically, $\Lambda$ is treated as a thermodynamic 
dynamic pressure $P$ via the relation $P=-\frac{\Lambda}{8\pi}=\frac{3}{8\pi \ell^2}$ in \cite{kastor09}, 
consistent with the observation in \cite{sy06} that the conjugate thermodynamic variable is proportional 
to a volume. It naturally further implies that the BH mass should be treated as  total 
gravitational enthalpy via the relation $M=H\equiv U+PV$  rather than internal energy $U$. Then the first 
law of BH thermodynamics read off
\begin{eqnarray}
dM &=& T dS+V dP + \Phi dQ+\Omega dJ ~.\label{dmgs}
\end{eqnarray}
where the thermodynamic volume can be defined as
\begin{eqnarray}
V_{t} &=& \left(\frac{\partial M}{\partial P}\right)_{S,Q,J}  ~.\label{vvw}
\end{eqnarray}
Where $\Phi$ is the electric potential measured at infinity. 
This also further suggests that one can write the BH equation of state as $P=P(V,T)$.

Again the naive geometric volume  for spherically symmetric Schwarzschild BH can be defined as 
in terms of BH event horizon \cite{bk74}:
\begin{eqnarray}
V_{+} &=& V_{s}=\frac{4}{3}\pi r_{+}^3  ~.\label{vsc}
\end{eqnarray}
It has been proposed that the thermodynamic volume $V_{t}$ in general is not equal to the naive 
geometric volume $V_{g}$ \cite{cvetic11}. 

However, the idea for constructing a VDW BH \cite{mann15} builds based on the classical work of 
$P-V$ criticality of RN-AdS BH by Kubiz\v{n}\'{a}k and Mann \cite{david12}.  Where the authors first showed that 
the thermodynamic properties of the charged AdS BH exhibits a number of similarities with the VDWs liquid/gas system.
Thus, it was natural to wonder if a BH with an equation of state identical to the VDWs fluid exists.

Therefore it is also natural to investigate here these above described features for recently 
discovered VDW BH \cite{mann15} which is analogous to the VDWs fluid. We prove that the thermodynamic volume is 
greater than the naive geometric volume for this BH. We also show that the Smarr-Gibbs-Duhem relation is satisfied. 
For generalized Smarr formula with zero cosmological constant for RN BH and KN BH one can see the work of 
Banerjee et al. \cite{rabin} and  the generalized mass formula with non-zero cosmological constant one 
must see another interesting work by Kastor et al. \cite{kastor09}.

We further investigate that the thermodynamic  stability of such BH by computing the specific heat and we observe 
that VDW BH undergoes no phase transition at all due to its own characteristics. The second order 
phase transition occurs at the negative value of the horizon thus it is \emph{unphysical}.  

In the first section, we have described the basic features of the VDW BH and we have also calculated the
specific heat which determins the local thermodynamic stability. In the second section, we have computed the logarithmic 
correction to the BH entropy for this class of BHs.

\section{VDW BH:}
The metric of a static, spherically symmetric  VDW  BH \cite{mann15} is given by
\begin{eqnarray}
ds^2=-{\cal B}(r)dt^{2}+\frac{dr^{2}}{{\cal B}(r)}+ r^{2}\left(d\theta^{2}+
\sin^{2}\theta d\phi^{2}\right) ~.\label{sph}
\end{eqnarray}
where the function ${\cal B}(r)$ is defined by
$$
{\cal B}(r)= 2\pi a-\frac{2M}{r}+\frac{r^2}{\ell^2}\left(1+\frac{3}{2}\frac{b}{r} \right)-
\frac{3\pi a b^2}{r(2r+3b)}
$$
\begin{eqnarray}
-\frac{4\pi ab}{r}\ln\left( \frac{r}{b}+\frac{3}{2}\right)  ~.\label{mf}
\end{eqnarray}
where $a$ and $b$ are the VDW parameters. It may be noted that the parameter $a>0$ implying 
the attraction in between the molecules of the fluid and indicating spherical horizon topology of the 
VDW  BHs. The parameter $b$ measures their volume and it also must be positive definite i.e. $b>0$.

The metric is a solution of the Einstein field equations of the form: $G_{ab}+\Lambda g_{ab}=8\pi T_{ab}$. It is
indeed true that for sufficiently small pressures,  the energy-momentum tensor $T_{ab}$ satisfies all three weak 
energy condition, strong energy condition and dominant energy conditions. However, when the pressure is increasing
the energy density is decreasing at small radii and eventually becomes negative, consequently first violates the 
dominant energy condition and followed by violates the weak energy condition. Whereas the strong energy 
condition is not violated. 

For simplicity, we take $a=\frac{1}{2\pi}$ then the metric function reduces to the form:
$$
{\cal B}(r) = 1-\frac{2M}{r}+\frac{r^2}{\ell^2}\left(1+\frac{3}{2}\frac{b}{r} \right)-
\frac{3}{2}\frac{ b^2}{r(2r+3b)}
$$
\begin{eqnarray}
-2\frac{b}{r}\ln\left( \frac{r}{b}+\frac{3}{2}\right)  ~.\label{mf1}
\end{eqnarray}

Now the mass \footnote{It should be noted that computing the mass parameter for VDW BH has 
some problems. Specifically, the matter content does not meet the standard fall-off conditions and therefore normal 
methods for calculating the mass e.g. conformal completion, fail. The reason for calling the parameter $M$ the mass is 
simply because it satisfies the first law of thermodynamics \cite{mann15}.} of the BH could be expressed in terms of 
event horizon $r_{+}$ :
\begin{eqnarray}
M = \frac{r_{+}}{2}+\frac{r_{+}^2}{2 \ell^2}\left(r_{+}+\frac{3b}{2}\right)-\frac{3}{4}\frac{b^2}{2r_{+}+3b}
-b\ln\left( \frac{r_{+}}{b}+\frac{3}{2}\right)  ~.\label{mf2}
\end{eqnarray}

Penrose \cite{rp} in 1973 had made a statement that the total ADM (Arnowitt-Deser-Misner) mass $M$ of the 
Schwarzschild BH spacetime is connected to the area $A_{+}$  of BH event horizon as 
\begin{eqnarray}
M  &\geq&  \sqrt{\frac{A_{+}}{16\pi}} ~.\label{vpi}
\end{eqnarray}
which is sometimes called  \emph{Cosmic Censorship Inequality} or \emph{Cosmic Censorship Bound} \cite{gibb05}. 
This is a necessary condition for cosmic-censorship hypothesis \cite{rp} (See \cite{bray,bray1,jang,rg,gibb99}). 

We compute this inequality for this BH which is given by 
$$
M  \geq  \sqrt{\frac{A_{+}}{16\pi}}+\frac{A_{+}}{8\pi\ell^2} \left( \sqrt{\frac{A_{+}}{4\pi}} +\frac{3}{2} b\right)
-\frac{3}{4} \frac{b^2}{\left( \sqrt{\frac{A_{+}}{\pi}} +3b\right)} 
$$
\begin{eqnarray}
-b \ln\left( \sqrt{\frac{A_{+}}{4\pi b^2}} +\frac{3}{2}\right) ~.\label{vpi1}
\end{eqnarray}
It suggests the total mass in a given region of the spacetime is at least  $\sqrt{\frac{A_{+}}{16\pi}}$ (for Schwarzschild BH) 
and positive definite. The significance of this inequality is that it  describes the lower bound on the energy for 
any time-symmetric initial data set which  satisfied the Einstein equations with negative cosmological constant, and which 
is also coupled to a matter system that satisfied the dominant energy condition which possesses no naked
singularities. 

The surface gravity of the BH on the horizon $r=r_{+}$ is given by 
\begin{eqnarray}
{\kappa}_{+} = \frac{{\cal B}'(r)}{2}= \frac{1 +\frac{3 r_{+}(r_{+}+b)}{\ell^2}+\frac{ 3b^2 }{(2r_{+}+3b)^2}-
\frac{4b}{(2r_{+}+3b)}}{2r_{+}} ~.\label{sgv}
\end{eqnarray}
which is constant over the horizon indicates the zeroth law of thermodynamics is satisfied.

Consequently, the BH temperature on the horizon is given by 
\begin{eqnarray}
T_{+} &=&  \frac{1 +\frac{3 r_{+}(r_{+}+b)}{\ell^2}+\frac{ 3b^2 }{(2r_{+}+3b)^2}-
\frac{4b}{(2r_{+}+3b)}}{4 \pi r_{+}} ~.\label{tmv}
\end{eqnarray}

The BH area reads off
\begin{eqnarray}
 A_{+} &=& \int^{2\pi}_0\int^\pi_0 \sqrt{g_{\theta\theta}g_{\phi\phi}} d\theta d\phi =4\pi r_{+}^2
~.\label{area}
\end{eqnarray}

The BH entropy is computed on the horizon as  (in units in which $G=\hbar=c=1$)
\begin{eqnarray}
 S_{+} &=& \frac{{\cal A}_{+}}{4} =\pi r_{+}^2  ~.\label{etpvw}
\end{eqnarray}
Now the mass of the BH could be expressed as in terms of entropy $S_{+}$ and dynamic pressure $P$ :
$$
M =\frac{1}{2} \sqrt{\frac{S_{+}}{\pi}}-  \frac{3}{4} \frac{b^2}{2 \sqrt{\frac{S_{+}}{\pi}}+3b}
-b\ln\left( \frac{1}{b}\sqrt{\frac{S_{+}}{\pi}}+\frac{3}{2}\right)
$$
\begin{eqnarray}
+\frac{4}{3} S_{+} P \left[ \sqrt{\frac{S_{+}}{\pi}}+\frac{3}{2}\right]
 ~.\label{mf3}
\end{eqnarray}

The most important parameter in the gravitational thermodynamics can be defined as
\begin{eqnarray}
H &=& M(S,P) = U+PV    ~.\label{enth}
\end{eqnarray}
Where $H$ is so called enthalpy of the gravitational system and $U$ is its  internal energy. If the natural candidate
$H$ is a function of $S$ and $P$, the first law of gravitational thermodynamics yield
\begin{eqnarray}
dH &=& T_{+}dS_{+}+V_{+} dP    ~.\label{enth1}
\end{eqnarray}
Thus we get the usual thermodynamical parameters like temperature, volume etc. as
$$
T_{+} = \left(\frac{\partial H}{\partial S_{+}}\right)_{P}=\frac{1}{4\sqrt{\pi S_{+}}}+
$$
$$
\frac{3}{4} \frac{b^2}{\sqrt{\pi S_{+}}(2 \sqrt{\frac{S_{+}}{\pi}}+3b)^2}
-\frac{1}{2 \sqrt{\pi S_{+}} \left( \frac{1}{b}\sqrt{\frac{S_{+}}{\pi}}+\frac{3}{2}\right) }
$$
\begin{eqnarray}
 +2 P  \left( \sqrt{\frac{S_{+}}{\pi}}+b\right)     ~.\label{tem}
\end{eqnarray}
and
\begin{eqnarray}
V_{t} &=& \left(\frac{\partial H}{\partial P}\right)_{S} =\frac{4}{3} S_{+}  \left( \sqrt{\frac{S_{+}}{\pi}}+\frac{3b}{2}\right)
 ~.\label{vol}
\end{eqnarray}
Thus it is defined as a thermodynamic volume  and the naive geometric volume  $V_{g}$ which
can be rewritten as
\begin{eqnarray}
V_{g} &=& \frac{4}{3} S_{+} \sqrt{\frac{S_{+}}{\pi}}  ~.\label{vol1}
\end{eqnarray}
It is clearly evident from above derived relations that
\begin{eqnarray}
V_{t} > V_{g}  ~.\label{vol2}
\end{eqnarray}
which was first pointed out in \cite{cvetic11} for higher dimensional rotating BH.
This is one of the key point of our investigation. The significance of this result is related to the reverse 
isoperimetric inequality. In particular, that this implies that the inequality is strict. The initial conjecture for the 
reverse isoperimetric inequality was first formulated in \cite{cvetic11} and in case of Super-Entropic BHs were discussed 
in \cite{mannprl}.

Now using dimensional analysis, the Smarr-Gibbs-Duhem relation could be derived as
\begin{eqnarray}
H &=& M = 2(TS-PV)   ~.\label{sgdv}
\end{eqnarray}
and the internal energy reads off
\begin{eqnarray}
U &=& 2 TS-3PV   ~.\label{sgd}
\end{eqnarray}
Finally, the Gibbs free energy is calculated to be
\begin{eqnarray}
G &=& H-TS=TS-2PV   ~.\label{gfe}
\end{eqnarray}

Now let us calculate the local thermodynamic stability for this BH. In order to determine the thermodynamic 
stability, one must compute the specific heat  which is given by the well known formula:
\begin{eqnarray}
C_{+} &=& \left(\frac{\partial M}{\partial T_{+}} \right)= \frac{ \left(\frac{\partial M}{\partial r_{+}} \right)}
{ \left(\frac{\partial T_{+}}{\partial r_{+}} \right)}  ~.\label{cv}
\end{eqnarray}
The specific heat in this case becomes
\begin{eqnarray}
C_{+} = -2\pi r_{+}^2 \frac{\left[1 +\frac{3 r_{+}(r_{+}+b)}{\ell^2}- \frac{(8br_{+} +9b^2) }{(2r_{+}+3b)^2}\right ]}
{ \left[1-\frac{(32br_{+}^2+54b^2r_{+}+27b^3)}{(2r_{+}+3b)^3}-\frac{3 r_{+}^2}{\ell^2}\right]}.~\label{cv2}
\end{eqnarray}

Now we analyze the above expression of specific heat for different regime.

\emph{Case I:}
When 
\begin{eqnarray}
 1 +\frac{3 r_{+}(r_{+}+b)}{\ell^2} > \frac{(8br_{+} +9b^2) }{(2r_{+}+3b)^2} \,\, \mbox{and} \nonumber\\
 1 > \frac{(32br_{+}^2+54b^2r_{+}+27b^3)}{(2r_{+}+3b)^3}+\frac{3 r_{+}^2}{\ell^2}
\end{eqnarray}
the specific heat is negative i. e. $C_{+}<0$, which implies that the BH is thermodynamically unstable.

\emph{Case II}: When  
\begin{eqnarray}
 1 +\frac{3 r_{+}(r_{+}+b)}{\ell^2} < \frac{(8br_{+} +9b^2)}{(2r_{+}+3b)^2} \,\, \mbox{and} \nonumber\\
 1> \frac{(32br_{+}^2+54b^2r_{+}+27b^3)}{(2r_{+}+3b)^3}+\frac{3 r_{+}^2}{\ell^2}\\
 \mbox{or} \nonumber\\
1 +\frac{3 r_{+}(r_{+}+b)}{\ell^2} > \frac{(8br_{+} +9b^2) }{(2r_{+}+3b)^2} \,\, \mbox{and}\,\, \nonumber\\
1 < \frac{(32br_{+}^2+54b^2r_{+}+27b^3)}{(2r_{+}+3b)^3}+\frac{3 r_{+}^2}{\ell^2}
\end{eqnarray}
the specific heat is positive  i. e. $C_{+}>0$, which indicates that  the BH is thermodynamically stable.

\begin{figure}[h]
\begin{center}
\subfigure[]{
\includegraphics[width=2.1in,angle=0]{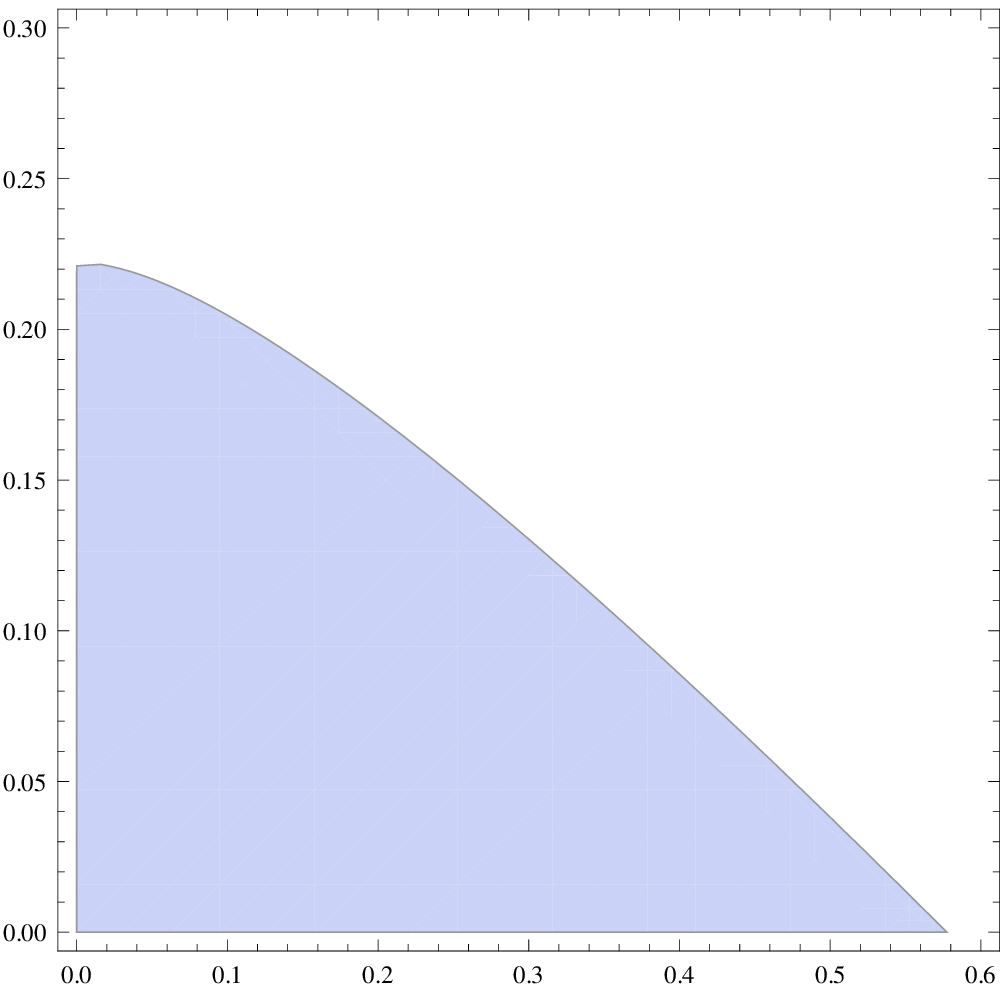}} 
\subfigure[]{
 \includegraphics[width=2.1in,angle=0]{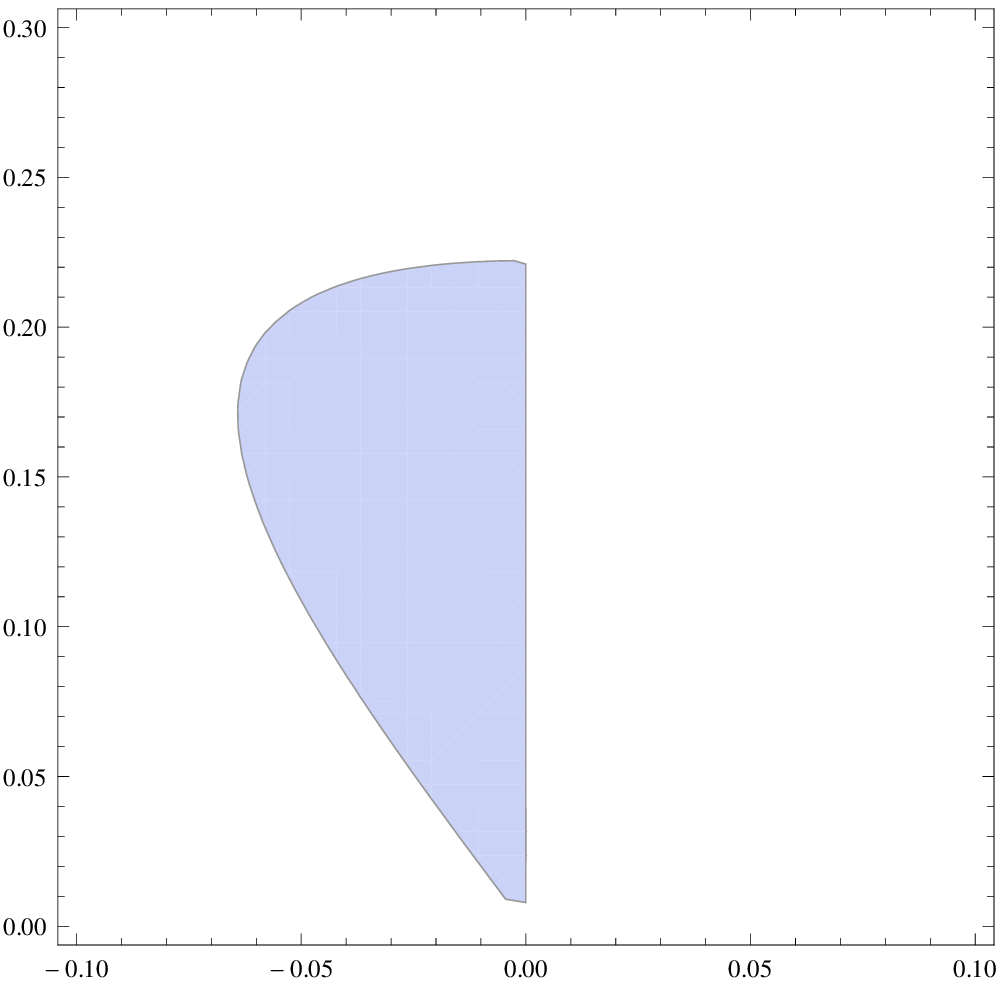}}
 \subfigure[]{
 \includegraphics[width=2.1in,angle=0]{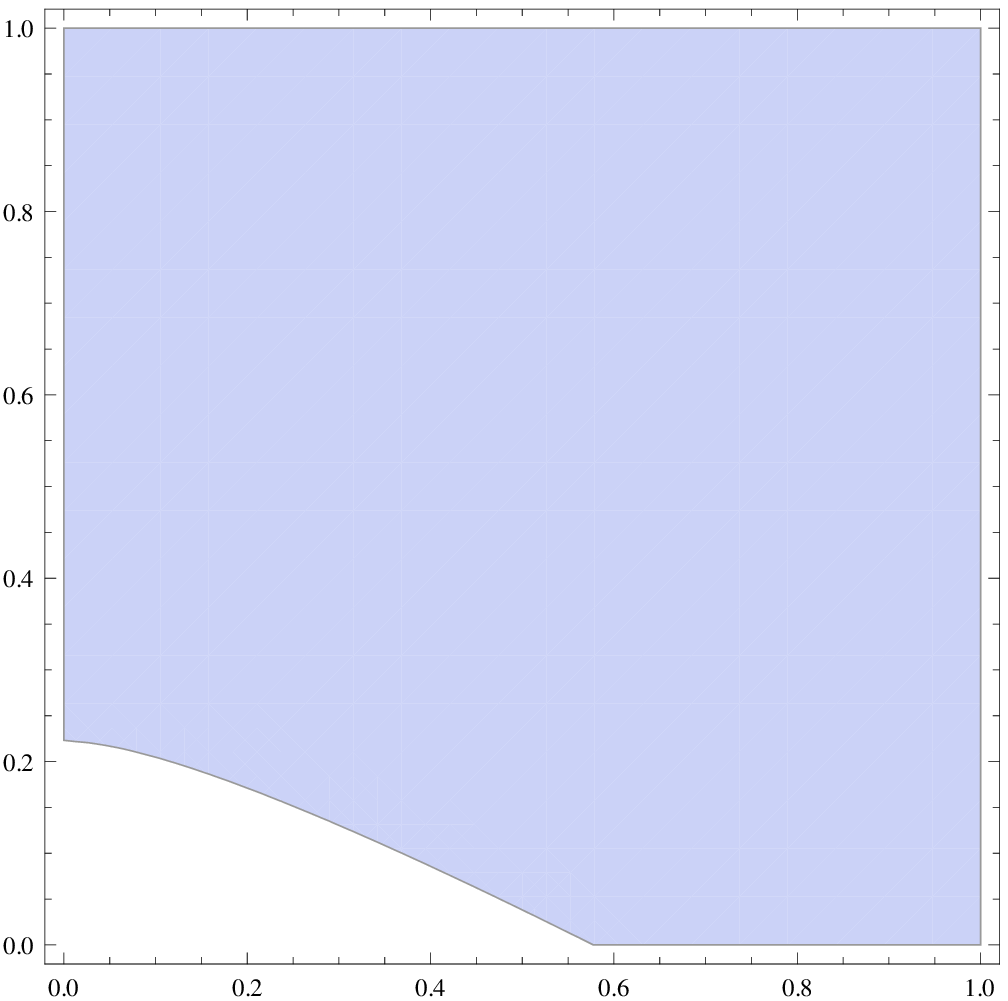}}
  \caption{\label{fg1}\textit{The inequality for specific heat of the Case-I is plotted in Fig. 1(a), Case-II is plotted in Fig. 1(b)
 and Fig. 1(c)}. Along the abcissa we have taken the value of horizon radius $r_{+}$ and along the ordinate we have taken the value 
 of $b$. We have set $\ell=1$. }
\end{center}
\end{figure}

\emph{Case III}: When 
\begin{eqnarray}
 1 &=& \frac{(32br_{+}^2+54b^2r_{+}+27b^3)}{(2r_{+}+3b)^3}+\frac{3 r_{+}^2}{\ell^2} 
\end{eqnarray}
the specific heat $C_{+}$ blows up that signals a second order phase transition for such BHs. Unfortunately, this 
phase transition occurs at the negative value of the horizon radius. Therefore it is unphysical. Thus we can only say that VDW 
BH does not possess any kind of second order phase transition. This is an interesting property of this BH. It could be observed 
from the Fig. \ref{fg2}. In Fig. \ref{fg1}, we have drawn  the region of validity of the inequality for the specific heat of 
different cases. 

\section{Logarithmic Corrections to the Entropy of VDW BH:}
In this section, we shall compute the logarithmic corrections to BH entropy for VDW's BH.  Assuming the  thermodynamic 
system is stable when its \emph{specific heat is positive definite} that implies the corresponding 
canonical ensemble is also thermodynamically stable. To compute it, we follow the work of Das et al. \cite{psm}. 
Recently, in \cite{mir} the effects of thermal fluctuations in case of charged ADS BH has been studied.

The partition function \cite{haw77} can be defined as
\begin{eqnarray}
 Z_{+}(\beta_{+}) &=& \int_{0}^{\infty} \rho_{+}(E_{+}) e^{-\beta_{+} E_{+}} dE_{+}  ~. \label{vpf}
\end{eqnarray}
where $T_{+}=\frac{1}{\beta_{+}}$ is the temperature of the ${\cal H}^{+}$. We have set Boltzman constant
$k_{B}=1$.

The density of states may be written as an inverse Laplace transformation of
the partition function:
\begin{eqnarray}
 \rho_{+}(E_{+}) &=& \frac{1}{2 \pi i}\int_{c-i\infty}^{c+i\infty} Z_{+}(\beta_{+})
 e^{\beta_{+} E_{+}} d\beta_{+} \\
 &=& \frac{1}{2 \pi i}\int_{c-i\infty}^{c+i\infty}
 e^{S_{+}(\beta_{+})} d{\beta_{+}}~. \label{vif}
\end{eqnarray}
where $c$ is a real constant and 
\begin{eqnarray}
S_{+} &=&  \ln Z_{+} +\beta_{+} E_{+}  ~. \label{vents}
\end{eqnarray}
is the entropy of the system near its equilibrium.

Near the equilibrium and at the inverse temperature $\beta_{+}=\beta_{0, +}$, we could expand the entropy
function as
\begin{eqnarray}
S_{+}(\beta_{+}) &=&  S_{0, +}+\frac{1}{2} (\beta_{+}-\beta_{0, +})^2 S_{0, +}'' + ...  ~.\label{sbs0}
\end{eqnarray}
where, $S_{0, +}: =S_{+}(\beta_{0, +})$ and $S_{0, +}''=\frac{\partial^2 S_{+}}{\partial \beta_{+}^2}$
at $\beta_{+}=\beta_{0, +}$.

Substituting Eq. (\ref{sbs0}) in Eq. (\ref{vpf}), we find
\begin{eqnarray}
\rho_{+}(E_{+}) &=& \frac{e^{S_{0, +}}}{2 \pi i}\int_{c-i\infty}^{c+i\infty}
e^\frac{\left(\beta_{+}-\beta_{0, +}\right)^2 S_{0, +}''}{2} d{\beta_{+}}~.\label{rhoe}
\end{eqnarray}
Let us define $\beta_{+}-\beta_{0, +} =i x_{+}$ and choose $c=\beta_{0, +}$, $x_{+}$ is a real variable 
and computing a contour integration one obtains
\begin{eqnarray}
 \rho_{+}(E_{+}) &=& \frac{e^{S_{0, +}}}{\sqrt{2 \pi S_{0, +}''}}~.\label{ps0}
\end{eqnarray}
The logarithm of the $\rho_{+}(E_{+})$ gives the corrected entropy of the system:
\begin{eqnarray}
{\cal S}_{+}:  &=&  \ln \rho_{+} ={\cal S}_{0, +}-\frac{1}{2} \ln S_{0, +}''+ ... ~. \label{srho}
\end{eqnarray}

Using the concept of statistical thermal fluctuations of mean value  energy \cite{sommer} for any thermodynamical 
system and it is  determined by the thermal heat capacity $C_{+}$ which may be defined as
\begin{eqnarray}
 C_{+} & \equiv & \frac{\partial <E_{+}>}{\partial T_{+}}\mid_{T_{0, +}} \\
 &=& \frac{1}{T_{+}^2}\left[\frac{1}{Z_{+}}\frac{\partial^2 Z_{+}}{\partial \beta_{+}^2}
 \mid_{\beta_{+}=\beta_{0, +}}-\frac{1}{Z_{+}^2}(\frac{\partial Z_{+}}{\partial \beta_{+}})^2
 \mid_{\beta_{+}=\beta_{0, +}} \right]\nonumber \\
 &=& \frac{S_{0, +}''}{T_{+}^2}~.\label{engsh}
\end{eqnarray}
where, the mean value of energy \cite{sommer} is 
\begin{eqnarray}
 <E_{+}> =-\frac{\partial}{\partial \beta_{+}}\ln Z_{+} \mid_{\beta_{+}=\beta_{0, +}} =
 -\frac{1}{Z_{+}}\frac{\partial Z_{+}}{\partial \beta_{+}}\mid_{\beta_{+}=\beta_{0, +}}
 ~.\label{engpf}
\end{eqnarray}

Therefore one obtains the leading order corrections to the BH entropy is
\begin{eqnarray}
{\cal S}_{+}  &=&  \ln \rho_{+} ={\cal S}_{0, +}-\frac{1}{2} \ln (C_{+} T_{+}^2)+...  ~. \label{canoetp}
\end{eqnarray}
The formula is meaningful when the  specific heat is positive definite in the logarithm term  thus we 
impose the condition $|C_{+}|>0$. Then the  Eq. (\ref{canoetp}) becomes
\begin{eqnarray}
{\cal S}_{+}  &=&  \ln \rho_{+} ={\cal S}_{0, +}-\frac{1}{2} \ln \left|C_{+} T_{+}^2\right|+...  ~ \label{ec}
\end{eqnarray}

Now apply this formula for VDW system and one obtains the logarithm correction to the BH entropy as 
$$
{\cal S}_{+}^{c} = \pi r_{+}^2-
$$
\begin{eqnarray}
\frac{1}{2} \ln \left|\frac{ 
\left[1 +\frac{3 r_{+}(r_{+}+b)}{\ell^2}-\frac{(8br_{+} +9b^2) }{(2r_{+}+3b)^2} \right]^3 }
{8\pi \left[1-\frac{(32br_{+}^2+54b^2r_{+}+27b^3)}{(2r_{+}+3b)^3}-
\frac{3 r_{+}^2}{\ell^2}\right]}  \right|+...  ~ \label{ec1}
\end{eqnarray}
It follows from the above calculation that this is the corrected microcanonical entropy due to quantum thermal 
fluctuations around the equilibrium. It should be mentioned that ${\cal S}_{+}^{c}$ is valid in the domain 
when 
\begin{eqnarray}
 1 +\frac{3 r_{+}(r_{+}+b)}{\ell^2} > \frac{(8br_{+} +9b^2) }{(2r_{+}+3b)^2} \,\, \mbox{and} \nonumber\\
 1 > \frac{(32br_{+}^2+54b^2r_{+}+27b^3)}{(2r_{+}+3b)^3}+\frac{3 r_{+}^2}{\ell^2}
\end{eqnarray}

Another way we can also derive the logarithmic correction to the BH entropy using exact entropy function 
${\cal S}_{+}(\beta_{+})=c\beta_{+}+\frac{d}{\beta_{+}}$ ($c,d$ are constants) followed by conformal 
field theory (CFT) \cite{carlip1,psm}. It could be take more general form as 
${\cal S}_{+}(\beta_{+}) = c \beta_{+}^{m}+\frac{d}{\beta_{+}^{n}}$ ($m, n, c, d>0$). The special case  
when $m=n=1$ and it is due to the CFT. After some algebra (more details could be found in \cite{psm,pp}) one 
obtains the $S_{0, +}''=T_{+}^2 S_{0, +}$,  then the leading order corrections to the generic BH entropy formula 
should be 
\begin{eqnarray}
{\cal S}_{+}  &=&  \ln \rho_{+} ={\cal S}_{0, +}-\frac{1}{2} \ln \left| T_{+}^2 S_{0, +} \right|+...  ~ \label{ve1}
\end{eqnarray}
After substituting the values of ${\cal S}_{0, +}=\pi r_{+}^2$ and $T_{+}$, one could find the logarithm correction 
to the BH entropy for VDW BH:
$$
{\cal S}_{+}^{cft}  = \pi r_{+}^2-
$$
\begin{eqnarray}
\frac{1}{2} \ln 
\left|\frac{\left[1 +\frac{3 r_{+}(r_{+}+b)}{\ell^2}-\frac{(8br_{+} +9b^2) }{(2r_{+}+3b)^2} \right]^2 }
{16\pi}\right|+...  ~ \label{ec2}
\end{eqnarray}
This equation is slightly numerically different from Eq. (\ref{ve1}) but it is quite interesting because of it does 
not depends upon the specific heat. It should  be noted that  ${\cal S}_{+}^{cft}$ is valid in the regime 
when 
\begin{eqnarray}
 1 +\frac{3 r_{+}(r_{+}+b)}{\ell^2} > \frac{(8br_{+} +9b^2) }{(2r_{+}+3b)^2}
\end{eqnarray}

Now we compare these two entropy corrected formula. When we have not taken the logarithmic correction, we have found the ratio
\begin{eqnarray}
\frac{{\cal S}_{+}^c}{{\cal S}_{+}^{cft}}  &=& 1  ~ \label{rt}
\end{eqnarray}
that means the two entropy is equal and it is expected. Now interesting thing happens when we have taken the logarithmic 
correction and the ratio should read off
$$
\frac{{\cal S}_{+}^c}{{\cal S}_{+}^{cft}} =
$$
\begin{eqnarray}
\frac{1-\frac{1}{2\pi r_{+}^2} \ln \left|\frac{ 
\left[1 +\frac{3 r_{+}(r_{+}+b)}{\ell^2}-\frac{(8br_{+} +9b^2) }{(2r_{+}+3b)^2} \right]^3 }
{8\pi \left[1-\frac{(32br_{+}^2+54b^2r_{+}+27b^3)}{(2r_{+}+3b)^3} -
\frac{3 r_{+}^2}{\ell^2}\right]}  \right|+...}
{1- \frac{1}{2\pi r_{+}^2} \ln 
\left|\frac{\left[1 +\frac{3 r_{+}(r_{+}+b)}{\ell^2}-\frac{(8br_{+} +9b^2) }{(2r_{+}+3b)^2} \right]^2 }
{16\pi}\right|+...}  ~ \label{rt1}
\end{eqnarray}
Up to the author's knowledge, it is very difficult to say which is greater or less. The domain of validity is discussed 
earlier. Only, we can say the ratio is strictly function of event horizon radius, VDW parameters and 
AdS radius i.e. $f(r_{+},a,b,\ell)$ because we have chosen the value of $a=\frac{1}{2\pi}$ earlier.

It should be emphasized that due to logarithmic correction, it seems plausible that the corrected entropy 
could turn out to be \emph{negative}\footnote{Although the meaning of negative entropy is somewhat unclear. However 
it appears, for example, in higher curvature theories of gravity. The dominant perspective on the topic seems to be to regard
negative entropy as unphysical. From the perspective that the entropy counts microstates, negative entropy would correspond 
fractional microstates, which has dubious meaning. Thus, many people tend to excise the negative entropy solutions from their 
analysis. However often times negative entropy solutions have no other pathological behaviour (e.g. curvature singularities), 
so they seem perfectly valid as classical solutions of the equation of motion. More work would be needed to really understand 
what is going on in these cases. }. Since, the expressions of logarithmic corrections in our case are complicated, so there may 
be a  possibility that some interesting feature comes into playing a key role in this topic.

\begin{figure}[h]
 \begin{center}
 \subfigure[ ]{
 \includegraphics[width=2.1in,angle=0]{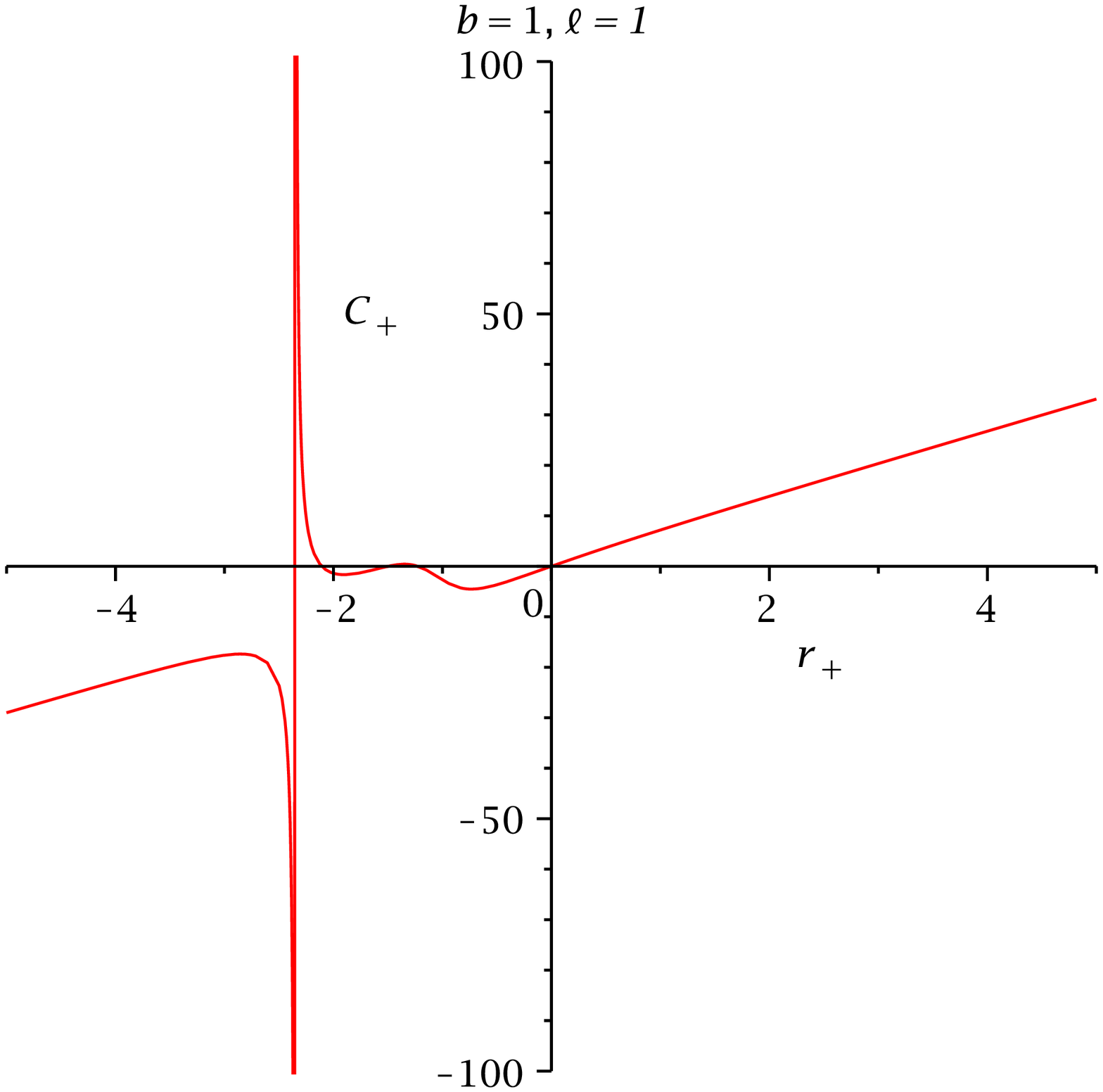}}
 \subfigure[ ]{
 \includegraphics[width=2.1in,angle=0]{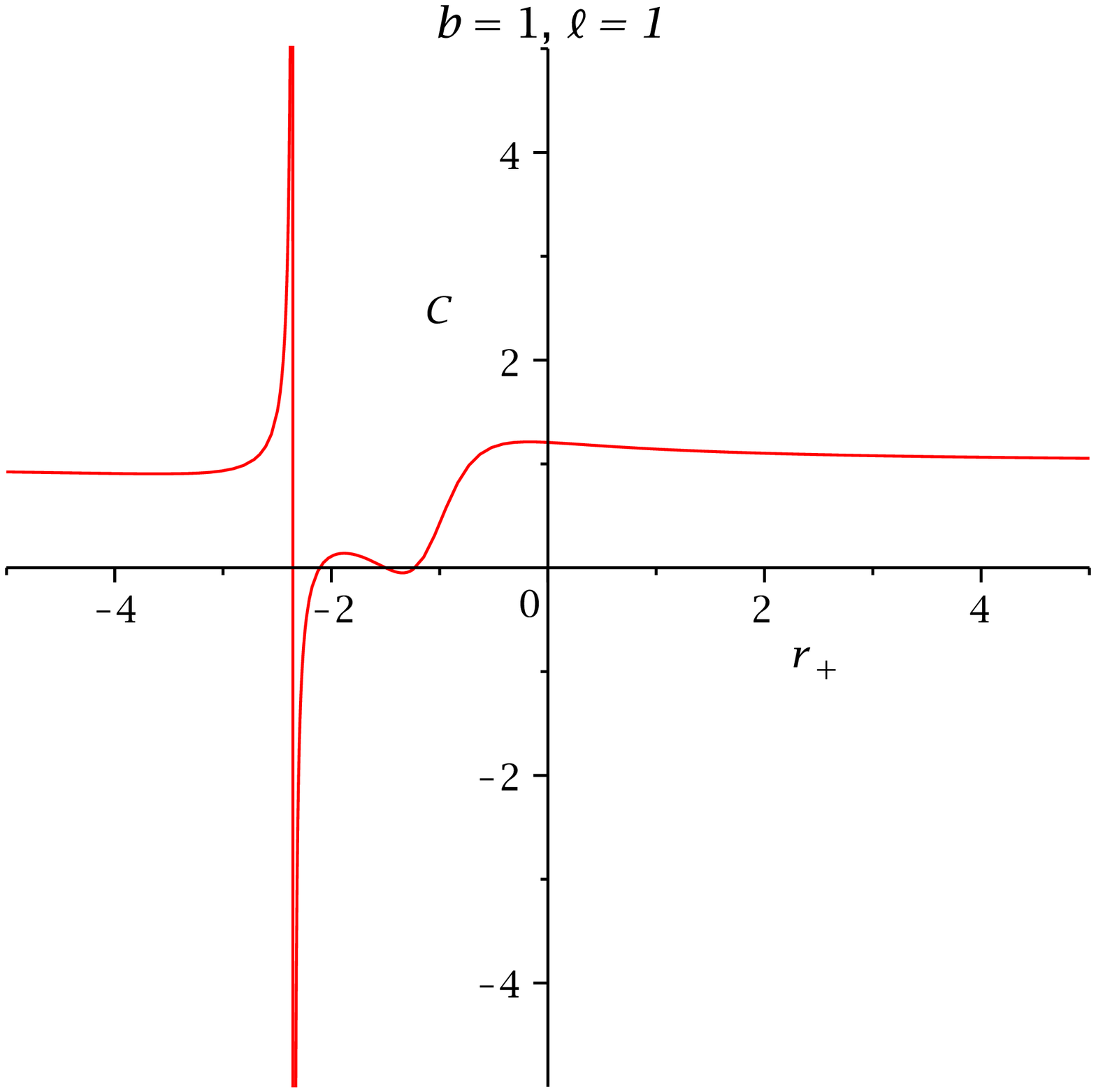}}
 \caption{\label{fg2}\textit{ In this figure, we have plotted the variation of specific heat with horizon radius for the values 
 $b=1$ and $\ell=1$. }}
\end{center}
\end{figure}

In Fig. \ref{fg2}, we have plotted the uncorrected and corrected specific heat with the horizon radius. 
From the figure one can speculated that the phase transition occurs at the  negative value of the event horizon
radius. 

Another way we can see the effect of logarithmic corrections by computing the corrected specific heat via the 
following relation and taking input from the above corrections then we find
$$
C = T_{+} \left(\frac{\partial S_{+}^c}{\partial T_{+}} \right)=-r_{+} 
\frac{\left[1 +\frac{3 r_{+}(r_{+}+b)}{\ell^2}- \frac{(8br_{+} +9b^2) }{(2r_{+}+3b)^2}\right ]}
{\left[1-\frac{(32br_{+}^2+54b^2r_{+}+27b^3)}{(2r_{+}+3b)^3}-\frac{3 r_{+}^2}{\ell^2}\right]} 
$$
$$
\times \left(\frac{\partial S_{+}^c}{\partial r_{+}} \right)
$$
where,
$$
\left(\frac{\partial S_{+}^c}{\partial r_{+}} \right) =2\pi r_{+}-\frac{3}{2}
\frac{\left[\frac{3}{\ell^2}(2r_{+}+b)+\frac{(16br_{+} +12b^2) }{(2r_{+}+3b)^3}\right]}
{\left[1 +\frac{3 r_{+}(r_{+}+b)}{\ell^2}- \frac{(8br_{+} +9b^2) }{(2r_{+}+3b)^2}\right ]}-
$$
\begin{eqnarray}
\frac{1}{2}\frac{\left[6\frac{r_{+}^2}{\ell^2}-\frac{(64br_{+}^2 +24b^2r_{+}) }{(2r_{+}+3b)^4}\right]}
{\left[1 +\frac{3 r_{+}(r_{+}+b)}{\ell^2}- \frac{(8br_{+} +9b^2) }{(2r_{+}+3b)^2}\right ]} +... \label{cv3}
\end{eqnarray}
The plot of the corrected specific heat could be seen  from the Fig. 2(b) and the uncorrected specific heat could 
be seen from the Fig. 2(a). From both the graph, it is clear that the ``phase transition'' occurs at negative 
horizon radius. This indicates that  the `phase transition''  is \emph{unphysical}. The only admissable 
values of $r$ should be positive (See, for example, the metric function which has terms like $\frac{1}{r}$ in it. These 
terms blow up at $r=0$ and asymptote to $\pm \infty$ to the left/right of the origin. This is an artefact of the curvature 
singularity that is at the origin. Furthermore, due to the presence of the logarithmic term, one can not restrict to just 
the negative values of $r$: the negative $r$ portion of the metric function terminates at some finite value of $r$, 
and hence would describe a compact space with no asymptotic region.) Thus, this result indicates that the \emph{VDW BH 
does not undergo a phase transition at all}, which is interesting in its own right.

\section{Discussion:}
Let us summarize the results. In the context of extended phase space, when the negative cosmological constant behaves
like a pressure and  volume be its thermodynamically conjugate variable then the gravitational mass could be 
described as total gravitational enthalpy rather than the energy. We have analyzed these features for recently discovered 
spherically symmetric VDW BH which satisfy all three: weak, strong and dominant energy conditions 
under small pressure. We computed the gravitational enthalpy of the said BH. We also computed the thermodynamic 
volume and proved that this volume is greater than the naive geometric volume. Furthermore, we have analyzed the
stablity of this BH. By evaluating the specific heat  we observed that  the VDW BH does not possess any 
kind of second order phase transition.  Smarr-Gibbs-Duhem relation is also satisfied for this BH.

Moreover, we derived the \emph{Cosmic-Censorship Inequality} for this BH.
Finally, we computed the \emph{logarithm correction} to the entropy for this BH due to the statistical quantum 
fluctuations around the  thermal equilibrium. These logarithmic corrected BH entropy formula due to statistical 
thermal fluctuations may have important application in Suskind's Holographic principle \cite{leo}. 

\section*{Acknowledgements:}

The author is grateful to the anonymous Referee for various nice suggestions and for which the manuscript 
enhanced substantially.

\end{document}